\documentclass[]{aa}  
\usepackage{graphicx, txfonts, natbib, hyperref}
\usepackage{amstext}

\begin{document}

   \title{Is beryllium ultra-depletion in solar-type stars linked 
          to the presence of a white dwarf companion?}

   \subtitle{}

   \author{S. Desidera\inst{1}
          \and
          V. D'Orazi\inst{1,2,3}
          \and
          M. Lugaro\inst{4,3}
          }

   \institute{INAF Osservatorio Astornomico di Padova, 
   vicolo dell'Osservatorio 5, I-35122, Padova, Italy\\
   \email{silvano.desidera@oapd.inaf.it}
  \and 
   Department of Physics and Astronomy, Macquarie University, Balaclava Rd, NSW 2109, Australia
   \and 
   Monash Centre for Astrophysics,  School of Physics and Astronomy, Monash University,  VIC 3800,  Australia
   \and
   Konkoly Observatory, Hungarian Academy of Sciences, PO Box 67, H-1525 Budapest, Hungary\\
   }
   
  \date{Received ; accepted }

 
  \abstract
   {Abundance studies of solar-type stars revealed a small fraction of objects
    with extreme depletion of beryllium.}
   {We investigate the possible link between the beryllium depletion
   and the presence of companions.}
   {The classical methods (radial velocity, astrometry, imaging) used to search 
    for binary companions were exploited. We also performed a
chemical 
    analysis to identify binaries by the alteration in abundances that is produced 
    by the accretion of material lost by a former evolved companion.}
   {We found that all the four previously investigated stars that were found to be ultra--depleted in Be are binaries. In two cases the companion is a 
    white dwarf, and in the other two cases the companion might be a white dwarf
    or a main-sequence star. One new barium star was identified.}
   {We speculate that the interaction with the white dwarf progenitor
    caused an alteration in the abundance pattern of the star,
which resulted 
    in severe
    beryllium depletion. Possible mechanisms such as thermohaline mixing, 
    episodic accretion, and rotational mixing are discussed. 
    We also briefly discuss predictions for validating this scenario. }


   \maketitle
%

\section{Introduction}

The determination of light-element abundances (Li, Be, and B) provides 
fundamental information on stellar interiors. Thanks to their fragile nature,
these three elements can be used to probe the mixing processes 
beneath the stellar surface, where they can be destroyed primarily by 
proton-capture reactions. Temperatures at which these phenomena occur are 
approximately 2.5 million K for Li and 4 million K for Be and B.
The spectral 
lines of B reside only in the far UV, which prevents gathering B abundances 
without going above the atmosphere. On the other hand, lithium is remarkably easier to observe (its 
resonance doublet occurs in the red portion of the spectrum at 6708 \AA), 
but because there are multiple sites for the creation and destruction of Li, 
the interpretation of Li abundances is sometimes complicated (see e.g. \citealt{kar14} 
and references therein). 
Beryllium provides a clearer picture because it has only one mechanism of production: 
spallation reactions in the vicinity of supernovae or in the ambient interstellar gas. 
The resonance lines of Be~{\sc ii} are found at 3130.421 and 3131.065 \AA, which are still 
observable with ground-based telescopes. For this reason, a plethora 
of observational projects have been commenced in the past decades to ascertain the Be content in 
different types of stars and stellar populations (old, metal-poor stars, globular clusters, 
solar-type stars and open clusters, including stars with exo-planets; see e.g.
\citealt{boesgard02}; \citealt{boesgard04}; \citealt{santos04}; \citealt{pasquini07}; 
\citealt{smiljanic09}, \citeyear{smiljanic10}, \citeyear{smiljanic11}). 

Recently, \citet{takeda11} presented Be abundances for a sample of 118 solar analogues (along with 87 FGK 
stars) and found that, although most of the Sun-like stars share a Be content similar to the Sun (roughly 
$\pm$ 0.2 dex within the solar value), four stars exhibit significant Be depletion by more than a factor 
of 100. They noted that these stars belong to the group with the slowest rotational velocities, where 
differential rotation between the core and the envelope may be higher and promote enhanced mixing leading 
to Be depletion (\citealt{bouvier08}). However, it is not straightforward to explain why this process 
should only occur in a tiny fraction of the relatively slowly
rotating Sun-like stars and not not in the 
large majority of these objects, or in other words, why it is not ubiquitous.

One of the four G dwarfs with very high beryllium depletion
identified by \citet{takeda11}, \object{HIP 64150},  was recently shown to have a white dwarf
(hereafter WD) companion at a projected separation of about 18 AU
\citep{crepp13}.
Motivated by this discovery, we examine the multiplicity of the four
Be-depleted G dwarfs to determine whether binarity is a unique case for HIP 64150
or a common feature of these stars.
We base our analysis on the standard techniques used to identify stellar
companions (e.g. radial velocity variability) 
and, more specifically for WD companions, on indirect evidence
of mass transfer from a former AGB star, seen as anomalies in their
chemical abundances.

\section{Binarity}
\label{s:bin}

\subsection{HIP 64150 = HD 114174}

As mentioned above, \object{HIP 64150} has a WD companion discovered 
as part of the TRENDS project (AO follow-up of stars with long-period RV trends)
by \citet{crepp13}.
The WD companion lies at a projected separation of about 18 AU
and the dynamical lower limit to its mass is  $0.26 M_{\odot}$.
\citet{crepp13} derived from their JHK photometry
an effective temperature around 8000~K and cooling age of about 3 Gyr
for the WD.
\citet{matthews14} estimated a significantly lower effective temperature and 
longer cooling age including their L'- and M-band photometry.
Regardeless of these discrepancies, the long cooling age is compatible with the slow
stellar rotation and low level of activity of the unevolved companion star.

\subsection{HIP 17336 = HD 23052}

\object{HIP 17336} is a spectroscopic and astrometric binary 
\citep[][ and Hipparcos orbital solution]{scarfe12}. The orbit has a period of 2.8 years 
and an eccentricity of 0.125.
The detection of the astrometric signature of the binary orbit from Hipparcos
data allowed \citet{scarfe12} to solve the inclination ambiguity 
from RV and infer a true companion mass of between 0.56 e 0.62 $M_{\odot}$, compatible 
with a late-K or early-M dwarf or a WD. 
In the latter case, considering its current separation, 
the system should have been evolved through 
the common-envelope phase. 
If the companion is a main-sequence star, the expected magnitude difference is about 3.8 
in V and 1.8 in K$_s$.
Therefore, the case of a main sequence companion can be easily confirmed or ruled
out with near-infrared observations \citep[see e.g. ][]{mazeh02}.
Galex far- and near-UV (FUV and NUV) and fluxes do not provide evidence for the presence of a hot WD.

\subsection{HIP 32673 = HD 49178}

This star is reported to have astrometric acceleration from Hipparcos.
\cite{nordstrom04} reported a possible variability (probability of constant RV
5\% from two-epoch observations).
To further evaluate the binarity of the object, we retrieved 12 RV measurements from
the ELODIE\footnote{\url{http://atlas.obs-hp.fr/elodie}} and 
SOPHIE\footnote{\url{http://atlas.obs-hp.fr/sophie/}} archive.
The four ELODIE spectra 
were taken in 2004-2005 and the eight SOPHIE spectra in late 2006.
While the time information is partially masked in the SOPHIE data archive, preventing
an accurate orbit determination, an RV regular trend of 1.6 km/s over the baseline of
the observations, ELODIE data also exhibit highly significant variations, and when
the \cite{nordstrom04} RV is taken into account, the peak-to-valley variations are of
15 km/s. A tentative periodicity of the order of 2.5-3 yr is deduced from  ELODIE and SOPHIE RVs.
The cross-correlation function (CCF) profiles from ELODIE and SOPHIE
indicate a single-line SB.
GALEX FUV and NUV fluxes do not provide evidence for the presence of a hot WD.

\subsection{HIP 75676 = HD 138004}

This star shows astrometric acceleration (seven-parameter solution) in Hipparcos.
Two-epoch RV observations from \cite{nordstrom04} show that this is an RV variable (rms RV 1.9 km/s, baseline 354d).
Furthemore, \cite{riddle15} reported that the star is a spectroscopic binary according to private communication by D. Latham.
Therefore,  this star probably also has a companion with a period of a few years.
From the signature of pollution by an AGB star that we found in Sect.~\ref{s:abu}, we conclude
that this companion must be a white dwarf. 
No additional companions were detected in the AO survey by \cite{metchev09}, 
however, there appears to also be a wide third companion at 40 
arcsec projected separation, the K dwarf BD+43 2500B.
This latter was recently resolved into a 0.4 arcsec close visual binary by \cite{riddle15}.

\section{Abundance analysis}
\label{s:abu}

Along with searching for the dynamical signature from a companion and the direct search
through imaging observations, 
an alternative way to infer the presence of a WD companion is to search
for abundance anomalies that are linked to the accretion of material
produced during the AGB phase, enriched in C and elements produced by the $slow\text{}$ neutron-capture process
\citep[the $s$ process, ][]{busso99}.
Barium stars represent the classical example of this process (\citealt{jorissen98}). 

\subsection{Data and analysis procedure}

\begin{center}
\begin{figure}
\includegraphics[width=9cm]{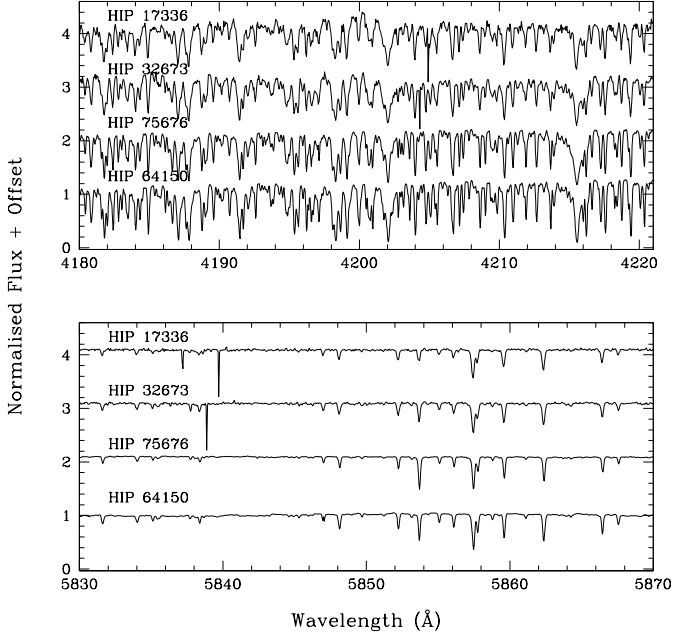}
\caption{Example of the spectra for our sample stars in two different wavelength regions.}\label{f:spectra}
\end{figure}
\end{center}

To derive the chemical abundances, we exploited spectra from
the High Resolution Echelle Spectrograph (HIRES, \citealt{vogt94}) that are available from the 
Keck archive\footnote{\url{http://www2.keck.hawaii.edu/koa/public/koa.php}} for stars HIP 64150 and HIP 75676. 
HIP 64150 was observed on June 9, 1997 under program N10H (PI: J. Lissauer)  employing the 0.6 arcsec slit, which results in a spectral resolving power of about R$\sim$55000. The wavelength coverage ranges from 
3750 to 6200 $\AA$. The signal-to-noise (S/N) ratio (per pixel) is about 200 at $\sim$ 5850 $\AA$. 
For HIP 75676 the spectrum was acquired during the observing program N03H (PI: J. Butler) on June 16, 2003; 
a 0.9 arcsec slit was adopted, which implies a resolution of R$\sim$45000
(spectral coverage 3750-6200 $\AA$). The S/N ratio around the Ba~{\sc ii} line at 5853 $\AA$~is 300.
For the other two sample stars, HIP 32673 and HIP 17336, we used 
ELODIE spectra available from the archive\footnote{\url{http://atlas.obs-hp.fr/elodie/}}. The spectrograph, which 
was mounted at the Cassegrain focus of the 1.93m telescope at the Observatoire de Haute-Provence (CNRS), 
covers a 3000 $\AA$~range (3850-6800 $\AA$) with a spectral resolution of R=42000.
Observations were carried out on February 17, 2004 (PI: T. Mazeh) and on February 18, 2005 (PI: T. Mazeh) 
for HIP 32673 and HIP 17336, respectively. The S/N ratios are around 70 and 50 at 5850 $\AA$. 
In Fig.~\ref{f:spectra} we show a portion of the spectra for the four sample stars in the 
wavelength regions 4180$-$4220 $\AA$ (upper panel) and 5830$-$5870 $\AA$ (lower panel).

\begin{center}
\begin{figure}
\includegraphics[width=9cm]{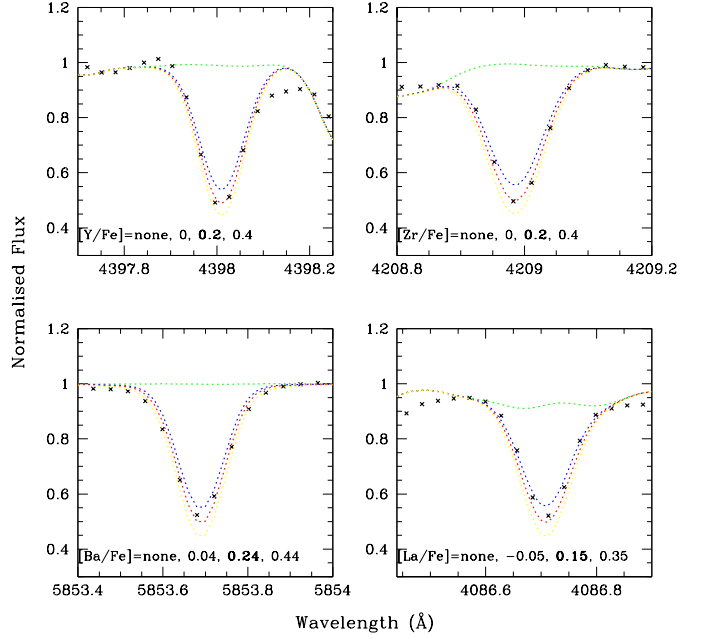}
\caption{Spectral synthesis for star HIP 64150 for the Y~{\sc ii}, Zr~{\sc ii}, Ba~{\sc ii,} and La~{\sc ii} 
lines.} \label{f:synth}
\end{figure}
\end{center}

Abundance analysis was carried out via spectral synthesis with MOOG (\citealt{sneden73}, 2011 version) and the Kurucz 
grid of model atmospheres with no overshooting (\citealt{ck03}). 
For $s$-process element abundances, we adopted the same line lists as in \cite{dorazi12}, to which we refer for details on atomic 
oscillator strengths, hyperfine structure information, isotopic shifts, and adopted ratios, and derived abundances 
for the Sun and Arcturus. Here we briefly recall that a single-line treatment was assumed for the Y~{\sc ii} 
and Zr~{\sc ii} lines (4398 $\AA$ and 4209 $\AA$, respectively), while hyperfine structure and isotopic shifts 
were included for the Ba~{\sc ii} feature at 5853 $\AA$ and La~{\sc ii} at 4087 $\AA$. An example of the 
spectral synthesis procedure is shown in Fig.~\ref{f:synth} for star HIP 64150.

Carbon abundances were obtained by exploiting the CH molecular band at 4300 \AA. The line list comes from 
Plez (private communication). For consistency with our previous analysis and $s$-process element 
abundances, we assumed $T_{\rm eff, \odot}$=5770K, log~$g_{\odot}$=4.44 cm~s$^{-2}$, $\xi_{\odot}$=1.1 
km~s$^{-1}$, and log~n(Fe)$_{\odot}$=7.52. Our analysis results in a carbon solar abundance of 
A(C)$_{\odot}$=8.30, which is lower than typical literature values, such as A(C)=8.52$\pm$0.06 by 
\cite{grevesse98}. However, since our analysis is strictly differential with respect to the Sun, this 
does not affect our conclusions.

Stellar parameters ($T_{\rm eff}$, log$g$, microturbulence velocity $\xi$) and metallicity ([Fe/H]) were 
retrieved from 
the study by \cite{takeda07}, who published Li abundances for our four stars (within 
a total sample of 118 stars). We refer to \cite{takeda07} for an extensive discussion 
on the accuracy of the atmospheric parameters
derived trough the spectroscopic procedure (excitation/ionisation equilibrium condition and curve-of-growth 
matching). 
Uncertainties on the derived abundances were evaluated following the same approach as in 
\citeauthor{dorazi09} (\citeyear{dorazi09}, \citeyear{dorazi11}, \citeyear{dorazi12}). 
The total internal errors related to the best-fit determination and to the adopted set of stellar parameters
range from 0.13 to 0.15 dex, depending on the species, with Ba affected by larger uncertainties 
given the almost saturated behaviour of the line at 5853 $\AA$ ($\xi$ is by far the dominant source of error).
Following \cite{takeda07}, we adopted errors of 20 K, 0.05 dex, 0.10 kms$^{-1}$, 
and 0.02 dex for $T_{\rm eff}$, log$g$, $\xi$, and input metallicity [Fe/H], respectively.


\subsection{Results}

Our results are reported in Table~\ref{t:results}, where we list stellar parameters 
(from \citealt{takeda07}), $V$ magnitudes and $B-V$ colours, along with 
abundances for carbon and $s$-process elements.
We found a solar composition in terms of Y, Zr, Ba, and La for HIP 32673, with a  
mean value [$s$/Fe]=0.01$\pm$0.01 (rms=0.03).
The star HIP 17336 exhibits, on the other hand, a sub-solar $s$-process element content, being the average
[$s$/Fe]=$-$0.27$\pm$0.03; in this case 
we could not derive the La abundance because of a cosmic ray falling on the La line at 4087$\AA$.

Interestingly, we identified a new Ba star,  HIP 75676: this star is 
characterised by a high over-abundance of $s$-process elements, both first-peak Y and Zr and 
second-peak Ba and La, with a mean value of [$s$/Fe]=1.16$\pm$0.02.

Finally, HIP 64150 reflects a mild super-solar abundance pattern in terms of $s$-process elements
([$s$/Fe]=0.21$\pm$0.01).
This star has been previously analysed by \cite{ramirez09}, who derived 
 [Ba/Fe]=0.15$\pm$0.05; the small difference ($\Delta$=0.09$\pm$0.10), 
which is within the measurement uncertainties, can be explained by the difference in the 
microturbulence velocity since we have adopted the value of $\xi$=1.03 by \citealt{takeda07}, 
while \citealt{ramirez09} obtained $\xi$=1.14 kms$^{-1}$. 

\begin{center}
\begin{table*}
\caption{Stellar parameters, V magnitudes, BV colours, C and $s$-process element abundances.}\label{t:results}
\begin{tabular}{lcccccccccccr}
\hline\hline
Star      & Alt. name  & V & B-V & $T_{\rm eff}$ & log$g$ & [Fe/H]  & $\xi$        & [C/Fe] & [Y/Fe]   & [Zr/Fe] & [Ba/Fe] & [La/Fe]  \\
          &            &   &     &    (K)        &  (cm~s$^{-2}$)      &  (dex)  & (km s$^{-1}$) &  (dex)      &(dex)    &  (dex) &  (dex)   &  (dex)   \\
          &            &   &  5   &       &        &   &      &               &           &          &         &         \\
\hline
HIP 64150 & HD 114174  & 6.78 & 0.667 & 5747  & 4.45   & ~~~0.07 & 1.03 & 0.00 & ~~0.20   &  ~0.20    & ~~0.24   &  ~~~0.18 \\
HIP 75676 & HD 138004  & 7.48 & 0.676 & 5755  & 4.40   & $-$0.09 & 0.92 & 0.21 & ~~1.20   &  ~1.10    & ~~1.20   &  ~~~1.15  \\ 
HIP 32673 & HD 49178~~ & 8.07 & 0.677 & 5724  & 4.57   & ~~~0.06 & 0.95 & $-$0.05 & ~~0.00   &  ~0.00    & ~~0.05   &  $-$0.02   \\ 
HIP 17336 & HD 23052~~ & 7.07 & 0.659 & 5671  & 4.54   & $-$0.13 & 0.94 & $-$0.12 &$-$0.30   &$-$0.20    &$-$0.30   &  ~~.....    \\        
          &            &               &      &       &   &        &         &       &           &          &          &         \\             
\hline\hline
\end{tabular}
\end{table*}
\end{center}
\begin{center}
\begin{figure}
\includegraphics[width=9cm]{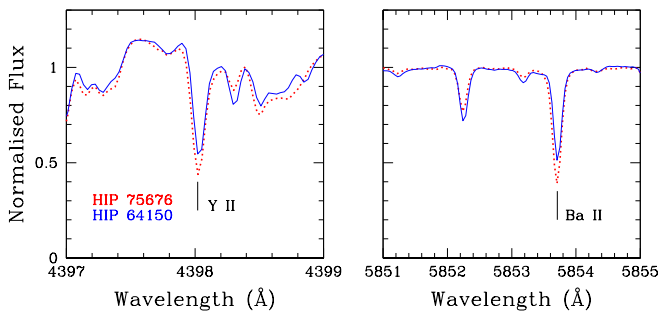}
\caption{Comparison of HIP 64150 (blue solid line) 
and HIP 75676 (red dotted line)} \label{f:comparison}
\end{figure}
\end{center}

\section{Discussion}

Our investigation of the multiplicity of the four ultra Be-depleted G dwarfs
identified by \citet{takeda11} shows that all these objects are binaries,
with companions with orbital period from a few years to several tens of years.
At least two of them have WD companions: one was shown by
direct detection, the other was deduced based on its $s$-process overabundance.
The remaining two cases are also compatible with a WD companion, especially 
HIP 17336, for which the dynamical mass of the companion is between 0.56 to 0.62
$M_{\odot}$.

\begin{center}
\begin{table*}
\caption{Binaries in the \citet{takeda11} sample and their characteristics.
SB: spectroscopic binaries, AB: astrometric binaries (with orbital solution), AA: stars
with astrometric acceleration, VB: visual binaries, EB: eclipsing binaries, 
Ba: barium stars (abundance anomalies indicating
a WD companion). In the semimajor axis column the asterisks indicate the projected separation in AU
as measured from imaging observation. HIP 112504 is a triple system. The moderately wide
visual companion and the close spectroscopic companion are listed as separated entries.
References: Cr13: \citet{crepp13}; Ma14: \citet{matthews14};
Sc12: \citet{scarfe12}; WDS: \citet{wds};   SB9: \citet{sb9}; HIP: \citet{hipparcos}; 
Go07: \citet{goldin07}; To14: \citet{tokovinin14}; To15: \citet{tokovinin15};
Gu05: \citet{guenther05}; Cr12: \citet{crepp12}
Ma12: \citet{malkov12}; No04: \citet{nordstrom04}.
}\label{t:binaries}
\begin{tabular}{lcccccccccccr}
\hline\hline
Star       & Alt. name  &  A(Be)    & Period     & sep    &  a        &  e & M1 & M2  & Type   & Reference  \\
           &            &           &            & arscec & AU        &    & $M_{\odot}$   & $M_{\odot}$    &        &            \\
           &            &           &            &        &           &    &    &     &        &            \\
\hline
HIP 64150 & HD 114174   & $<-0.88$  &   --       &  0.68  & 18.0$^{*}$ & --    & 1.05 & 0.54 & SB+VB    & Cr13, Ma14 \\ 
HIP 75676 & HD 138004   & $<-0.96$  &   --       &  --    &  --       &  --   & 0.96     &  --  & SB+AB+Ba & this paper \\ 
HIP 32673 & HD 49178    & $<-0.78$  & $\sim$1000d &  --    &   --     & --    & 1.00     &  --  & SB+AA    & this paper \\ 
HIP 17336 & HD 23052    & $<-0.85$  &  1030d      &   --   &   --     & 0.125 & 0.95 & 0.59 & SB+AB    & Sc12 \\ 
\hline\hline
HIP 7244   & HD  9472   &  1.23     &  --        & 2.8    & 97.0$^{*}$ &  --   & 0.98 & 0.11 &  VB     & WDS         \\  
HIP 7918   & HD  10307  &  1.25     & 7122d       & 0.7    &  8.9      & 0.10  & 1.08 & 0.43 &  SB+AA+VB  & SB9,HIP,WDS  \\  
HIP 8486   & HD  11131  &  1.33     & 2471d       &  --    &  4.0      & 0.45  & 0.97 & 0.42 &   AB    &   Go07   \\ 
HIP 19911  & HD  26990  &  1.33     & 5.68y       & 0.10   &  3.10     &  0.81    & 0.95 & 0.85 &  SB,AA,VB  &  HIP, To14, To15           \\ 
HIP 20441  & HD  27685  &  1.16     & 4623d       &  -- & 6.16      &  --    & 1.01 & 0.45 &  SB+AA      & HIP, To14            \\ 
HIP 20741  & HD  28099  &  1.25 &  --    & 0.43   & 19.8$^{*}$ &  --    & 1.08 & 0.16    &  SB+VB     &   Gu05  \\ 
HIP 25414  & HD  35073  &  1.25 &  --    &  --   &  --   &  --    & 1.02 & --    &  AA        &   HIP        \\ 
HIP 40118  & HD  68017  &  1.08 &  --    & 0.6   & 13.0  &  --    & 0.81 & 0.15  & SB+VB      &   Cr12       \\
HIP 41184  & HD  70516  &  1.70 &  --    & 21.0  & 790   & 0.94   & 1.02 & 0.81  & VB         &   To14, Ma12 \\ 
HIP 43557  & HD  75767  &  1.19 & 10.25d & --    &  0.1  & 0.10   & 1.00 & 0.30  &  SB+EB      & SB9          \\ 
HIP 49728  & HD  88084  &  1.30 &  --    & 2.3   & 79.3$^{*}$  &  --      & 1.00 & 0.47 &  VB+AB  &  To14     \\ 
HIP 78217  & HD 144061  &  1.15 &  --    & 2.30  &  68.1$^{*}$ &   --   & 0.95 & 0.48 & VB      & To14      \\ 
HIP 89912  & HD 168874  &  1.42 &  --    & 1.50  &  42.8$^{*}$ &   --   & 1.05 & 0.52 & SB+AA+VB   & To14 \\   
HIP 96402  & HD 184768  &  1.28 &  --    & 3.20  & 123.5$^{*}$ &   --   & 0.96 & 0.39 & VB      & To14       \\ 
HIP 104075 & HD 200746  &  1.32 & 91.91y & 0.48  &  21.2      &  0.47  & 1.09 & 0.64  & VB   & Ma12,To14  \\  
HIP 109110 & HD 209779  &  1.41 & 13.1yr & 0.082 &   2.9 &  -- & 1.04 & 0.62 &  SB+AA+VB  & To14            \\ 
HIP 112504 & HD 215696  &  1.37 &  --    & 2.71  &  91.8$^{*}$ &   --   & 1.02 &  0.75 &  VB     &  HIP, To14           \\
           &            &       & 420d   &  --   &    --       &   --  & 1.02 &   --   &  SB     &  To14           \\
HIP 115715 & HD 220821  &  1.22 &  --    & 0.41 &  15.0$^{*}$ &   --   & 0.88 & 0.65 &  SB+VB  &  WDS,To14,No04  \\  

\hline\hline
\end{tabular}
\end{table*}
\end{center}

We investigated the multiplicity of the remaining 114 solar analogues in the \cite{takeda11} 
sample in the SB9 \citep{sb9}, Hipparcos \citep{hipparcos}, and WDS catalogues \citep{wds}.
We found that several stars with normal Be abundance have stellar companions
with a variety of separations, encompassing the range of ultra Be-depleted stars (Table \ref{t:binaries}).
None of these companions, however, are WDs.
This indicates that even though a WD companion needs to be confirmed for two of our targets, the 
observed Be depletion must be linked not only to the presence of a stellar companion, but specifically 
to the case when the companion star already evolved throughout the red giant and AGB 
phases into a WD. The presence of a WD companion within a few AU implies some interaction between the 
G dwarf and the WD progenitor during its red giant or AGB phase
through a common-envelope phase and/or wind 
accretion, depending on the initial binary configuration. 
Accretion of mass from an AGB companion is the standard explanation 
for the C- and $s$-process-enhanced composition of Ba stars, as in the case of the Ba 
star found in our sample, and it is possible also for the mildly enriched star reported here. 
The C- and $s$-process enrichment is expected only in AGB stars of masses roughly above 1.5 M$_{\odot}$ 
because they experience the third dredge-up \citep{iben83}. The other two stars that show 
no $s$-process enhancements could also have experienced the same binary evolution path leading to Be depletion 
as the stars showing $s$-process enhancements, but with interaction and accretion from an AGB companion of 
initial mass lower than 1.5 M$_{\odot}$. 

It is expected that the Be abundance in the 
accreted material is close to zero, since the envelope of AGB stars reaches very deep in 
mass to hot layers and strong Be depletions are observed already in the previous 
red giant phase \citep{2014PASJ...66...91T}. 
However, accretion from an AGB star is not enough by itself to quantitatively explain 
the observed level of depletion. 
The mass of the convective surface layer of these solar analogues is of the order of 0.02 
M$_{\odot}$. It would be necessary to accrete 2 M$_{\odot}$ of material from an AGB star to achieve 
the ratio of initial to accreted material of less than 1/100 required to explain the Be depletions 
of two orders of magnitude. This is clearly not feasible.
A possibility is that the material that 
we see at the stellar surface represents pure AGB material. 
Evidence of the presence of pure accreted material may come from a systematic depletion in C in the two stars 
that do not show $s$-process enhancement. The companions 
would have experienced the first dredge-up during the red giant phase, which reduces C, but 
not the third dredge-up during the AGB, which increases C. However, the C depletion during the 
first dredge-up depends 
on the initial stellar mass and may be too small to be detected 
($<$ 0.1 dex for an initial mass 
of the WD $<$ 1.2 M$_{\odot}$). 
Furthermore, it is difficult to find a physical reason why the 
accreted material should not mix into the convective surface layer of these stars.
Another three possible mechanisms to explain the origin of extreme Be depletion in solar-type stars are 
thermohaline mixing, episodic accretion, and rotational mixing.

\subsection{Thermohaline mixing}

Accretion from a red giant or AGB companion produces a change in the mean molecular weight of the external 
layers of the star, since the accreted material is more rich in He than the material on the secondary. 
The change in mean molecular weight is of the order of less than one to a few percent, depending on 
the amount of material accreted and if the companion experienced only the first dredge-up or also 
the third dredge-up. In any case, it would favour the development of mixing processes such as 
thermohaline mixing, as investigated in the context of carbon-enhanced metal-poor (CEMP) 
stars, together with effect of gravitational settling \citep[e.g.][]{2008MNRAS.389.1828S}. 
Depending on the amount of mass accreted and the initial mass of the companion, it is 
qualitatively expected that more efficient thermohaline mixing has been at work to deplete 
Be in the stars considered here than in similar stars that did not accrete from an
evolved companion, but only a detailed model can confirm this quantitatively.

\subsection{Episodic accretion}

\citet{baraffe10} and \citet{viallet12} showed that  bursts of episodic accretion 
phenomena during the pre-MS phase have significant effects on the stellar
structure that imply severe destruction of Li and Be.
The accretion bursts considered in these studies have rates of the order
of $10^{-4}$ - $10^{-3}~M_{\odot}/yr$, which may occur only in rather
extreme conditions of proto-stellar clouds.
The mass-loss rate of AGB stars is between 
10$^{-7}$ and 10$^{-4}$ $M_{\odot}/yr$,  
and these high required accretion values are also well beyond our expectation for the 
wind-accretion scenario, even if we assumed that significantly higher mass-loss rates may be achieved 
on short timescales. Conversely, common-envelope evolution 
may be rather fast, although much more uncertain, allowing in principle high accretion rates.
It should also be taken into account that the accretors considered in 
\citet{baraffe10} and \citet{viallet12} are protostars at early evolutionary
 phases, very different from the mature solar-type stars onto which the accretion 
would occur at the end of the AGB phase of the WD companion progenitor.
In these cases lower accretion rates might be enough to trigger
changes in the internal structure capable of destroying Be.

\subsection{Rotational mixing}

The accretion of material lost by the WD progenitor also implies accretion of angular
momentum on the companion, altering the rotational evolution of the companion
\citep{1996MNRAS.279..180J}. Several objects with WD companions
showing excesses of rotation and activity for their ages have been identified, from the extreme
case of the barium-rich, ultra-fast rotator 2REJ0357+283  \citep{1996A&A...315L..19J}, to the intermediate 
case of HD 8049 \citep{zurlo13}, 
to the planet-host GJ 86, whose slightly enhanced activity was identified through the very old kinematic age
and abundance pattern \citep{desidera07}.
This variety is the natural product of intrinsic differences in the amounts of accreted material
and its angular momentum and of following evolution of the fast-rotating star, driven by
magnetized stellar wind in a similar way to fast-rotating young stars \citep{barnes03}.
As a result of the accretion and spin-up, differential rotation between the core and the 
envelope may develop in solar-type stars and induce mixing and Be depletion, 
similarly to the single slow-rotating scenario 
invoked by \citet{takeda11} and in the pre-main sequence fast rotators 
considered by \citet{viallet12}.
The fastest rotation occurs for stars that have recently interacted with the
WD progenitor. All our program stars are characterized by slow rotation. Therefore, we expect
that the interaction occurred $>1$ Gyr ago and the WD companions should have cooling ages
longer than this value.
The rotational mixing scenario is favoured by the recent discovery of another solar analogue with $s$-process-enhancements 
and Be depletion (by more than a factor of fifteen, i.e., milder than in the stars considered here) but 
showing a rotational velocity significantly higher than solar \citep{schirbel15}. 
This star probably represents a progenitor of the stars investigated here.

\section{Concluding remarks}

Our study showed that all the four stars that are ultra-depleted in Be identified
by \citet{takeda11} are members of moderately close binaries.
Several other Be-normal stars are binaries, in some case with
similar separation, indicating that binarity itself is not the decisive factor
causing the Be depletion.
Instead, in two cases, the visual binary HIP 64050 and the newly discovered 
barium star HIP 75676, the companion is a confirmed WD and, while available data
are not conclusive with respect to a main-sequence companion, 
a WD companion is also fully compatible for the other two cases
(HIP 17336 and HIP 32673).
We speculate that the interaction with the WD progenitor
caused alteration in the abundance pattern of the star that resulted in the severe
Be depletion. We discussed possible mechanisms such as thermohaline mixing, 
episodic accretion, and rotational mixing. 

To validate our scenario, a more detailed characterization of companions
of stars with depleted or normal Be abundances should be carried out
to determine their nature as WD or main-sequence stars.
The determination of the orbital parameters of the pairs including WD companion
will also be highly valuable, allowing us to constrain the original binary 
configuration and the amount of mass exchanges that occurred during the AGB phase.

A complementary path is to study the Be abundance of stars that are known to have
a WD companion close enough to have produced significant mass and angular momentum
accretion onto the companion.
Main-sequence barium stars are a natural example, as mass transfer events 
should have been occurred while the companion (now a WD) was on the AGB, 
and produced carbon and s-process elements.
If our scenario is correct, we expect that all barium stars are depleted 
in beryllium.

\begin{acknowledgements}
We thank the Haute-Provence Observatory for maintaining ELODIE and SOPHIE archives.
This research has made use of the Keck Observatory Archive (KOA), 
which is operated by the W. M. Keck Observatory and the NASA Exoplanet 
Science Institute (NExScI), under contract with the National Aeronautics 
and Space Administration.
This research has made use of the SIMBAD database,
operated at CDS, Strasbourg, France and of the Washington Double Star Catalog 
maintained at the U.S. Naval Observatory.
ML is a Momentum (Lend\"ulet-2014' Programme) project leader of the 
Hungarian Academy of Sciences.
\end{acknowledgements}


\bibliography{beryllium}

\begin{thebibliography}{48}
\expandafter\ifx\csname natexlab\endcsname\relax\def\natexlab#1{#1}\fi

\bibitem[{{Baraffe} \& {Chabrier}(2010)}]{baraffe10}
{Baraffe}, I. \& {Chabrier}, G. 2010, \aap, 521, A44

\bibitem[{{Barnes}(2003)}]{barnes03}
{Barnes}, S.~A. 2003, \apj, 586, 464

\bibitem[{{Boesgaard} \& {King}(2002)}]{boesgard02}
{Boesgaard}, A.~M. \& {King}, J.~R. 2002, \apj, 565, 587

\bibitem[{{Boesgaard} {et~al.}(2004){Boesgaard}, {McGrath}, {Lambert}, \&
  {Cunha}}]{boesgard04}
{Boesgaard}, A.~M., {McGrath}, E.~J., {Lambert}, D.~L., \& {Cunha}, K. 2004,
  \apj, 606, 306

\bibitem[{{Bouvier}(2008)}]{bouvier08}
{Bouvier}, J. 2008, \aap, 489, L53

\bibitem[{{Busso} {et~al.}(1999){Busso}, {Gallino}, \& {Wasserburg}}]{busso99}
{Busso}, M., {Gallino}, R., \& {Wasserburg}, G.~J. 1999, \araa, 37, 239

\bibitem[{{Castelli} \& {Kurucz}(2003)}]{ck03}
{Castelli}, F. \& {Kurucz}, R.~L. 2003, in IAU Symposium, Vol. 210, Modelling
  of Stellar Atmospheres, ed. N.~{Piskunov}, W.~W. {Weiss}, \& D.~F. {Gray},
  20P

\bibitem[{{Crepp} {et~al.}(2012){Crepp}, {Johnson}, {Howard}, {Marcy},
  {Fischer}, {Hillenbrand}, {Yantek}, {Delaney}, {Wright}, {Isaacson}, \&
  {Montet}}]{crepp12}
{Crepp}, J.~R., {Johnson}, J.~A., {Howard}, A.~W., {et~al.} 2012, \apj, 761, 39

\bibitem[{{Crepp} {et~al.}(2013){Crepp}, {Johnson}, {Howard}, {Marcy},
  {Gianninas}, {Kilic}, \& {Wright}}]{crepp13}
{Crepp}, J.~R., {Johnson}, J.~A., {Howard}, A.~W., {et~al.} 2013, \apj, 774, 1

\bibitem[{{Desidera} \& {Barbieri}(2007)}]{desidera07}
{Desidera}, S. \& {Barbieri}, M. 2007, \aap, 462, 345

\bibitem[{{D'Orazi} {et~al.}(2012){D'Orazi}, {Biazzo}, {Desidera}, {Covino},
  {Andrievsky}, \& {Gratton}}]{dorazi12}
{D'Orazi}, V., {Biazzo}, K., {Desidera}, S., {et~al.} 2012, \mnras, 423, 2789

\bibitem[{{D'Orazi} {et~al.}(2011){D'Orazi}, {Biazzo}, \& {Randich}}]{dorazi11}
{D'Orazi}, V., {Biazzo}, K., \& {Randich}, S. 2011, \aap, 526, A103

\bibitem[{{D'Orazi} {et~al.}(2009){D'Orazi}, {Magrini}, {Randich}, {Galli},
  {Busso}, \& {Sestito}}]{dorazi09}
{D'Orazi}, V., {Magrini}, L., {Randich}, S., {et~al.} 2009, \apjl, 693, L31

\bibitem[{{Goldin} \& {Makarov}(2007)}]{goldin07}
{Goldin}, A. \& {Makarov}, V.~V. 2007, \apjs, 173, 137

\bibitem[{{Grevesse} \& {Sauval}(1998)}]{grevesse98}
{Grevesse}, N. \& {Sauval}, A.~J. 1998, \ssr, 85, 161

\bibitem[{{Guenther} {et~al.}(2005){Guenther}, {Paulson}, {Cochran},
  {Patience}, {Hatzes}, \& {Macintosh}}]{guenther05}
{Guenther}, E.~W., {Paulson}, D.~B., {Cochran}, W.~D., {et~al.} 2005, \aap,
  442, 1031

\bibitem[{{Iben} \& {Renzini}(1983)}]{iben83}
{Iben}, Jr., I. \& {Renzini}, A. 1983, \araa, 21, 271

\bibitem[{{Jeffries} \& {Smalley}(1996)}]{1996A&A...315L..19J}
{Jeffries}, R.~D. \& {Smalley}, B. 1996, \aap, 315, L19

\bibitem[{{Jeffries} \& {Stevens}(1996)}]{1996MNRAS.279..180J}
{Jeffries}, R.~D. \& {Stevens}, I.~R. 1996, \mnras, 279, 180

\bibitem[{{Jorissen} {et~al.}(1998){Jorissen}, {Van Eck}, {Mayor}, \&
  {Udry}}]{jorissen98}
{Jorissen}, A., {Van Eck}, S., {Mayor}, M., \& {Udry}, S. 1998, \aap, 332, 877

\bibitem[{{Karakas} \& {Lattanzio}(2014)}]{kar14}
{Karakas}, A.~I. \& {Lattanzio}, J.~C. 2014, \pasa, 31, 30

\bibitem[{{Malkov} {et~al.}(2012){Malkov}, {Tamazian}, {Docobo}, \&
  {Chulkov}}]{malkov12}
{Malkov}, O.~Y., {Tamazian}, V.~S., {Docobo}, J.~A., \& {Chulkov}, D.~A. 2012,
  \aap, 546, A69

\bibitem[{{Mason} {et~al.}(2001){Mason}, {Wycoff}, {Hartkopf}, {Douglass}, \&
  {Worley}}]{wds}
{Mason}, B.~D., {Wycoff}, G.~L., {Hartkopf}, W.~I., {Douglass}, G.~G., \&
  {Worley}, C.~E. 2001, \aj, 122, 3466

\bibitem[{{Matthews} {et~al.}(2014){Matthews}, {Crepp}, {Skemer}, {Hinz},
  {Gianninas}, {Kilic}, {Skrutskie}, {Bailey}, {Defrere}, {Leisenring},
  {Esposito}, \& {Puglisi}}]{matthews14}
{Matthews}, C.~T., {Crepp}, J.~R., {Skemer}, A., {et~al.} 2014, \apjl, 783, L25

\bibitem[{{Mazeh} {et~al.}(2002){Mazeh}, {Prato}, {Simon}, {Goldberg},
  {Norman}, \& {Zucker}}]{mazeh02}
{Mazeh}, T., {Prato}, L., {Simon}, M., {et~al.} 2002, \apj, 564, 1007

\bibitem[{{Metchev} \& {Hillenbrand}(2009)}]{metchev09}
{Metchev}, S.~A. \& {Hillenbrand}, L.~A. 2009, \apjs, 181, 62

\bibitem[{{Nordstr{\"o}m} {et~al.}(2004){Nordstr{\"o}m}, {Mayor}, {Andersen},
  {Holmberg}, {Pont}, {J{\o}rgensen}, {Olsen}, {Udry}, \&
  {Mowlavi}}]{nordstrom04}
{Nordstr{\"o}m}, B., {Mayor}, M., {Andersen}, J., {et~al.} 2004, \aap, 418, 989

\bibitem[{{Pasquini} {et~al.}(2007){Pasquini}, {Bonifacio}, {Randich}, {Galli},
  {Gratton}, \& {Wolff}}]{pasquini07}
{Pasquini}, L., {Bonifacio}, P., {Randich}, S., {et~al.} 2007, \aap, 464, 601

\bibitem[{{Perryman} {et~al.}(1997){Perryman}, {Lindegren}, {Kovalevsky},
  {Hoeg}, {Bastian}, {Bernacca}, {Cr{\'e}z{\'e}}, {Donati}, {Grenon},
  {Grewing}, {van Leeuwen}, {van der Marel}, {Mignard}, {Murray}, {Le Poole},
  {Schrijver}, {Turon}, {Arenou}, {Froeschl{\'e}}, \& {Petersen}}]{hipparcos}
{Perryman}, M.~A.~C., {Lindegren}, L., {Kovalevsky}, J., {et~al.} 1997, \aap,
  323, L49

\bibitem[{{Pourbaix} {et~al.}(2004){Pourbaix}, {Tokovinin}, {Batten}, {Fekel},
  {Hartkopf}, {Levato}, {Morrell}, {Torres}, \& {Udry}}]{sb9}
{Pourbaix}, D., {Tokovinin}, A.~A., {Batten}, A.~H., {et~al.} 2004, \aap, 424,
  727

\bibitem[{{Ram{\'{\i}}rez} {et~al.}(2009){Ram{\'{\i}}rez}, {Mel{\'e}ndez}, \&
  {Asplund}}]{ramirez09}
{Ram{\'{\i}}rez}, I., {Mel{\'e}ndez}, J., \& {Asplund}, M. 2009, \aap, 508, L17

\bibitem[{{Riddle} {et~al.}(2015){Riddle}, {Tokovinin}, {Mason}, {Hartkopf},
  {Roberts}, {Baranec}, {Law}, {Bui}, {Burse}, {Das}, {Dekany}, {Kulkarni},
  {Punnadi}, {Ramaprakash}, \& {Tendulkar}}]{riddle15}
{Riddle}, R.~L., {Tokovinin}, A., {Mason}, B.~D., {et~al.} 2015, \apj, 799, 4

\bibitem[{{Santos} {et~al.}(2004){Santos}, {Israelian}, {Garc{\'{\i}}a
  L{\'o}pez}, {Mayor}, {Rebolo}, {Randich}, {Ecuvillon}, \& {Dom{\'{\i}}nguez
  Cerde{\~n}a}}]{santos04}
{Santos}, N.~C., {Israelian}, G., {Garc{\'{\i}}a L{\'o}pez}, R.~J., {et~al.}
  2004, \aap, 427, 1085

\bibitem[{{Scarfe} \& {Griffin}(2012)}]{scarfe12}
{Scarfe}, C.~D. \& {Griffin}, R.~F. 2012, \rmxaa, 48, 257

\bibitem[{{Schirbel} {et~al.}(2015){Schirbel}, {Mel\'endez}, {Karakas},
  {Ram\'irez}, {Castro}, {Faria}, {Lugaro}, {Asplund}, {Tucci Maia}, {Yong},
  {Howes}, \& {do Nascimento}}]{schirbel15}
{Schirbel}, L., {Mel\'endez}, J., {Karakas}, A.~I., {et~al.} 2015, \aap,
  submitted

\bibitem[{{Smiljanic} {et~al.}(2009){Smiljanic}, {Pasquini}, {Bonifacio},
  {Galli}, {Gratton}, {Randich}, \& {Wolff}}]{smiljanic09}
{Smiljanic}, R., {Pasquini}, L., {Bonifacio}, P., {et~al.} 2009, \aap, 499, 103

\bibitem[{{Smiljanic} {et~al.}(2010){Smiljanic}, {Pasquini}, {Charbonnel}, \&
  {Lagarde}}]{smiljanic10}
{Smiljanic}, R., {Pasquini}, L., {Charbonnel}, C., \& {Lagarde}, N. 2010, \aap,
  510, A50

\bibitem[{{Smiljanic} {et~al.}(2011){Smiljanic}, {Randich}, \&
  {Pasquini}}]{smiljanic11}
{Smiljanic}, R., {Randich}, S., \& {Pasquini}, L. 2011, \aap, 535, A75

\bibitem[{{Sneden}(1973)}]{sneden73}
{Sneden}, C.~A. 1973, PhD thesis, THE UNIVERSITY OF TEXAS AT AUSTIN.

\bibitem[{{Stancliffe} \& {Glebbeek}(2008)}]{2008MNRAS.389.1828S}
{Stancliffe}, R.~J. \& {Glebbeek}, E. 2008, \mnras, 389, 1828

\bibitem[{{Takeda} {et~al.}(2007){Takeda}, {Kawanomoto}, {Honda}, {Ando}, \&
  {Sakurai}}]{takeda07}
{Takeda}, Y., {Kawanomoto}, S., {Honda}, S., {Ando}, H., \& {Sakurai}, T. 2007,
  \aap, 468, 663

\bibitem[{{Takeda} \& {Tajitsu}(2014)}]{2014PASJ...66...91T}
{Takeda}, Y. \& {Tajitsu}, A. 2014, \pasj, 66, 91

\bibitem[{{Takeda} {et~al.}(2011){Takeda}, {Tajitsu}, {Honda}, {Kawanomoto},
  {Ando}, \& {Sakurai}}]{takeda11}
{Takeda}, Y., {Tajitsu}, A., {Honda}, S., {et~al.} 2011, \pasj, 63, 697

\bibitem[{{Tokovinin}(2014)}]{tokovinin14}
{Tokovinin}, A. 2014, \aj, 147, 86

\bibitem[{{Tokovinin} {et~al.}(2015){Tokovinin}, {Mason}, {Hartkopf}, {Mendez},
  \& {Horch}}]{tokovinin15}
{Tokovinin}, A., {Mason}, B.~D., {Hartkopf}, W.~I., {Mendez}, R.~A., \&
  {Horch}, E.~P. 2015, \aj, 150, 50

\bibitem[{{Viallet} \& {Baraffe}(2012)}]{viallet12}
{Viallet}, M. \& {Baraffe}, I. 2012, \aap, 546, A113

\bibitem[{{Vogt} {et~al.}(1994){Vogt}, {Allen}, {Bigelow}, {Bresee}, {Brown},
  {Cantrall}, {Conrad}, {Couture}, {Delaney}, {Epps}, {Hilyard}, {Hilyard},
  {Horn}, {Jern}, {Kanto}, {Keane}, {Kibrick}, {Lewis}, {Osborne},
  {Pardeilhan}, {Pfister}, {Ricketts}, {Robinson}, {Stover}, {Tucker}, {Ward},
  \& {Wei}}]{vogt94}
{Vogt}, S.~S., {Allen}, S.~L., {Bigelow}, B.~C., {et~al.} 1994, in Society of
  Photo-Optical Instrumentation Engineers (SPIE) Conference Series, Vol. 2198,
  Society of Photo-Optical Instrumentation Engineers (SPIE) Conference Series,
  ed. D.~L. {Crawford} \& E.~R. {Craine}, 362

\bibitem[{{Zurlo} {et~al.}(2013){Zurlo}, {Vigan}, {Hagelberg}, {Desidera},
  {Chauvin}, {Almenara}, {Biazzo}, {Bonnefoy}, {Carson}, {Covino}, {Delorme},
  {D'Orazi}, {Gratton}, {Mesa}, {Messina}, {Moutou}, {Segransan}, {Turatto},
  {Udry}, \& {Wildi}}]{zurlo13}
{Zurlo}, A., {Vigan}, A., {Hagelberg}, J., {et~al.} 2013, \aap, 554, A21

\end{thebibliography}

\end{document}